\documentclass[
 reprint,
 amsmath,amssymb,
 aps,
 superscriptaddress
]{revtex4-2}

\usepackage{graphicx}
\usepackage{gensymb}

\usepackage{lmodern}
\usepackage[T1]{fontenc}

\begin{document}
\author{Luis Ledezma}
\affiliation{Department of Electrical Engineering, California Institute of Technology, Pasadena, California 91125, USA.}
\affiliation{Jet Propulsion Laboratory, California Institute of Technology, Pasadena, California 91109, USA.}

\author{Arkadev Roy}
\author{Luis Costa}
\author{Ryoto Sekine}
\author{Robert Gray}
\author{Qiushi Guo}
\author{Rajveer Nehra}
\affiliation{Department of Electrical Engineering, California Institute of Technology, Pasadena, California 91125, USA.}

\author{Ryan M. Briggs}
\affiliation{Jet Propulsion Laboratory, California Institute of Technology, Pasadena, California 91109, USA.}

\author{Alireza Marandi} \email{marandi@caltech.edu}
\affiliation{Department of Electrical Engineering, California Institute of Technology, Pasadena, California 91125, USA.}

\newenvironment{backmatter}{%
  \fontsize{8\p@}{10\p@}\selectfont%
  \newcommand{\bmsection}[1]{\par\medskip\noindent{\fontsize{9\p@}{11\p@}\bfseries ##1.\enspace}}%
}{}


\title{Octave-spanning tunable parametric oscillation in nanophotonics}

\begin{abstract}
Widely-tunable coherent sources are desirable in nanophotonics for a multitude of applications ranging from communications to sensing. The mid-infrared spectral region (wavelengths beyond 2 $\mu$m) is particularly important for applications relying on molecular spectroscopy. Among tunable sources, optical parametric oscillators typically offer some of the broadest tuning ranges; however, their implementations in nanophotonics have been limited to narrow tuning ranges and only at visible and near-infrared wavelengths. Here, we surpass these limits in dispersion-engineered periodically-poled lithium niobate nanophotonics and demonstrate ultra-widely tunable optical parametric oscillators. With a pump wavelength near 1 $\mu$m, we generate output wavelengths tunable from 1.53 $\mu$m to 3.25 $\mu$m in a single chip with output powers as high as tens of milliwatts. Our results represent the first octave-spanning tunable source in nanophotonics extending into the mid-infrared which can be useful for numerous integrated photonic applications.
\end{abstract}

\maketitle


Widely-tunable coherent sources are vital for applications ranging from multi-channel optical communications \cite{willner2020} to lidar \cite{jiang2020}. Wide tunability in the mid-infrared spectral range is especially desirable due to the rich molecular responses at wavelengths longer than 2 $\mu$m \cite{gordon2022}. While it is possible to generate light at these wavelengths with semiconductor lasers \cite{yao2012}, the tuning ranges are typically narrow due to limited bandwidth of semiconductor gain \cite{shim2021, han2022, li2022, beeck2021}. Alternatively, optical parametric oscillators (OPOs) based on quadratic nonlinearity have been a prominent example of sources with flexible and broad tuning ranges which have commonly been realized using nonlinear crystals in bulky table-top setups \cite{dunnParametricGenerationTunable1999}. 

Previous efforts towards OPO miniaturization include using lithium niobate diffused waveguides with fiber feedback loops \cite{langrock2007}, semiconductor waveguides with Bragg mirrors deposited on the chip end-facets \cite{savanier2013}, and lithium niobate whispering-gallery microresonators \cite{LN_WGR}.
However, implementation of OPOs in nanophotonics with sub-wavelength modal confinement and low propagation losses is highly desirable because of opportunities for dense integration with other on-chip components, strong nonlinear interactions, and dispersion engineering \cite{wang2018e}.

Over the past decade, nanophotonic OPO were demonstrated in the near-infrared and visible ranges using materials with cubic ($\chi^{(3)}$) and quadratic ($\chi^{(2)}$) nonlinearities \cite{razzari2010, lu2019, lu2020, bruch2019, lu2021, mckenna2022}. However, the main advantages of table-top OPOs, namely wide tunability and mid-infrared coverage, have not yet accessed in nanophotonics. A noteworthy roadblock for this is the typical use of simple pump-resonant configurations in which all the interacting optical fields resonate simultaneously in a single resonator. This leads to ultra-low OPO thresholds at the expense of an over-constrained wavelength tunability. In contrast, OPOs with singly- or doubly-resonant configurations (i.e. with non-resonant pump) offer wide tunability and frequency stability \cite{dunnParametricGenerationTunable1999}. 

Here, we design and demonstrate ultra-widely tunable doubly-resonant OPOs in lithium niobate nanophotonics. This is achieved by combining dispersion engineering, precise design of the spectral response of the cavity, and quasi-phase matching. With a pump tunable over ~30 nm at around 1 um, we achieve wavelengths tunable from 1.53 $\mu$m to 3.25 $\mu$m from five OPOs on a single nanophotonic chip.

\begin{figure*}[ht]
\centering
\includegraphics[trim={0 0cm 0 0cm}, clip, width=0.99\linewidth]{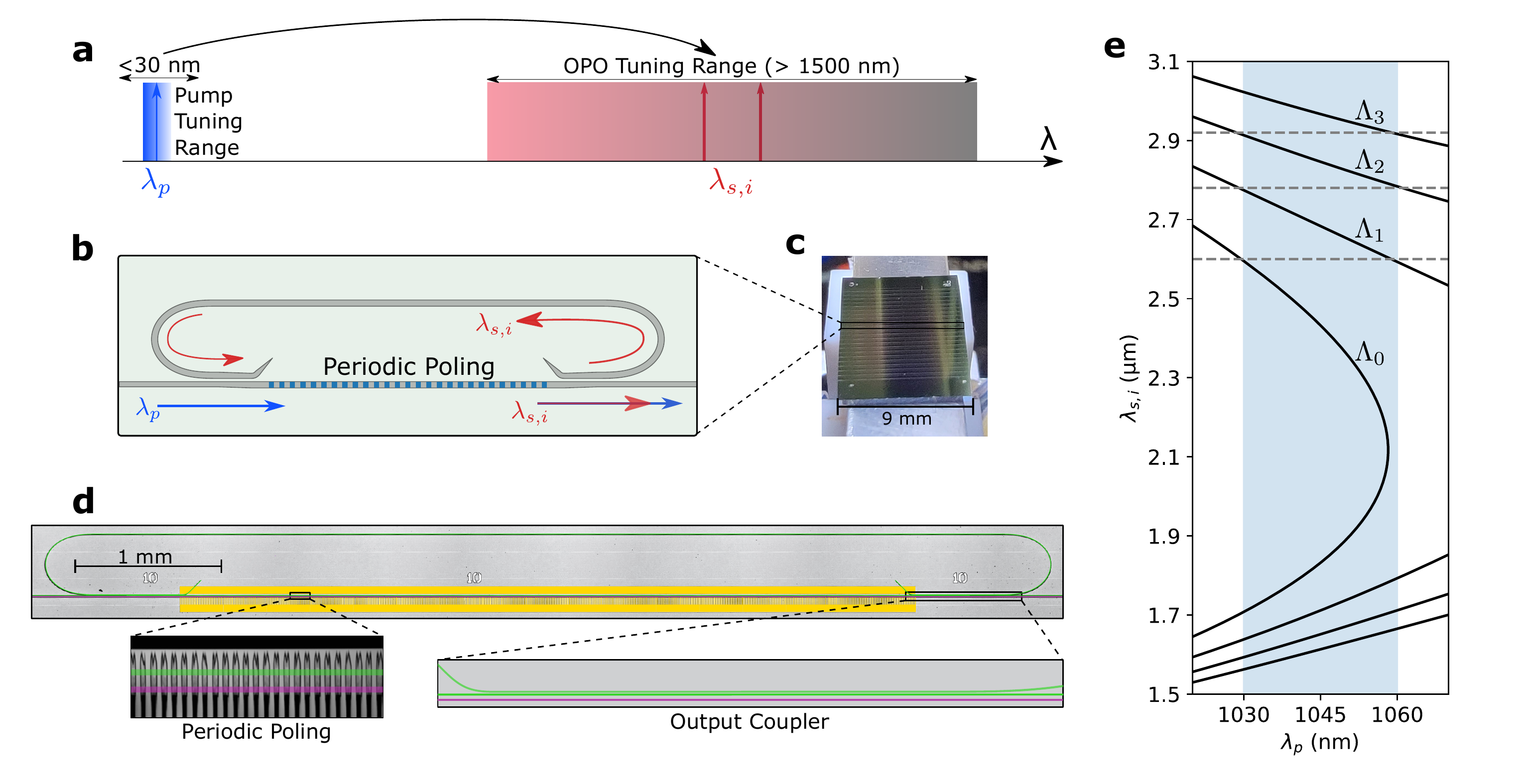}
\caption{
\textbf{Ultra-widely tunable optical parametric oscillators in nanophotonics.}
\textbf{a}, A narrowly tunable (<30 nm) pump around 1 um leads to an OPO signal and idler tuning range exceeding 1500 nm.
\textbf{b}, Schematic of the doubly-resonant parametric oscillator with a frequency-selective resonator that provides feedback only to the signal and idler while enabling continuous tuning of the pump.
\textbf{c}, Image of the chip highlighting the area occupied by a single OPO.
\textbf{d}, False-color optical microscope image of the OPO (green) and a straight waveguide (purple; used for calibration and phase-matching verification). Insets show a two-photon microscope image of the periodic poling and a close-up of the adiabatic output coupler.
\textbf{e}, Example of OPO tuning curves for four different poling periods $\Lambda_0 - \Lambda_3$. The dashed vertical lines and the blue stripe are to guide the eyes on how continuous tuning over an octave can be achieved with four poling periods and only 30 nm of pump tuning.
}
\end{figure*}

\section*{Results}

The tuning concept of the OPOs is illustrated in Fig. 1a, where more than 1500 nm of tuning around 2 $\mu$m for the signal and idler is obtained by tuning the pump wavelength around 1 $\mu$m by less than 30 nm. It is worth noting that such a tuning range for the pump is already available from integrated distributed Bragg reflector (DBR) semiconductor lasers \cite{verrinder2022}. Such magnification in tuning range from the pump towards the signal and idler (a factor of \textasciitilde12 in frequency units) is obtained through a dispersion-engineered quasi-phase matched OPO design with a spectrally-selective cavity as depicted in Fig. 1b. We use wavelength-selective couplers that allow the signal and idler wavelengths to resonate in the OPO cavity with a \textasciitilde 10-GHz free spectral range (FSR), while letting the pump go only through the poled waveguide section \cite{marandi2020}. This differs sharply from previously demonstrated fully resonant on-chip OPO designs in which the pump also needs to satisfy a resonant condition limiting their flexibility and tunability. A chip containing 16 OPOs is fabricated, as shown in Fig. 1c, where we have highlighted a single OPO, which is also displayed in the false-color optical microscope image of Fig. 1d.

The simulated tuning behavior of four OPOs with different poling periods are shown in Figure 1e (solid black lines). These are obtained from conservation of energy ($\omega_p = \omega_s + \omega_i$) and momentum ($k_p = k_s + k_i + 2\pi/\Lambda_{QPM}$), so they can be tailored by engineering the waveguide dispersion \cite{ledezma2022}. In particular, the signal tuning slope ($\partial \omega_s / \partial \omega_p$) is given by the ratio of group velocity differences $(1/v_i - 1/v_p)/(1/v_i - 1/v_s)$, while the gain-bandwidth is inversely proportional to $1/v_i - 1/v_s$. We have engineered the dispersion of the poled waveguide to balance these effects. As a result, a small change in the pump wavelength produces large changes in the output wavelengths while maintaining a predictable tuning curve  without substantial mode competition (see Extended Data Fig. \ref{XtraDesign}).

\begin{figure*}[t!]
\centering
\includegraphics[width=0.99\linewidth]{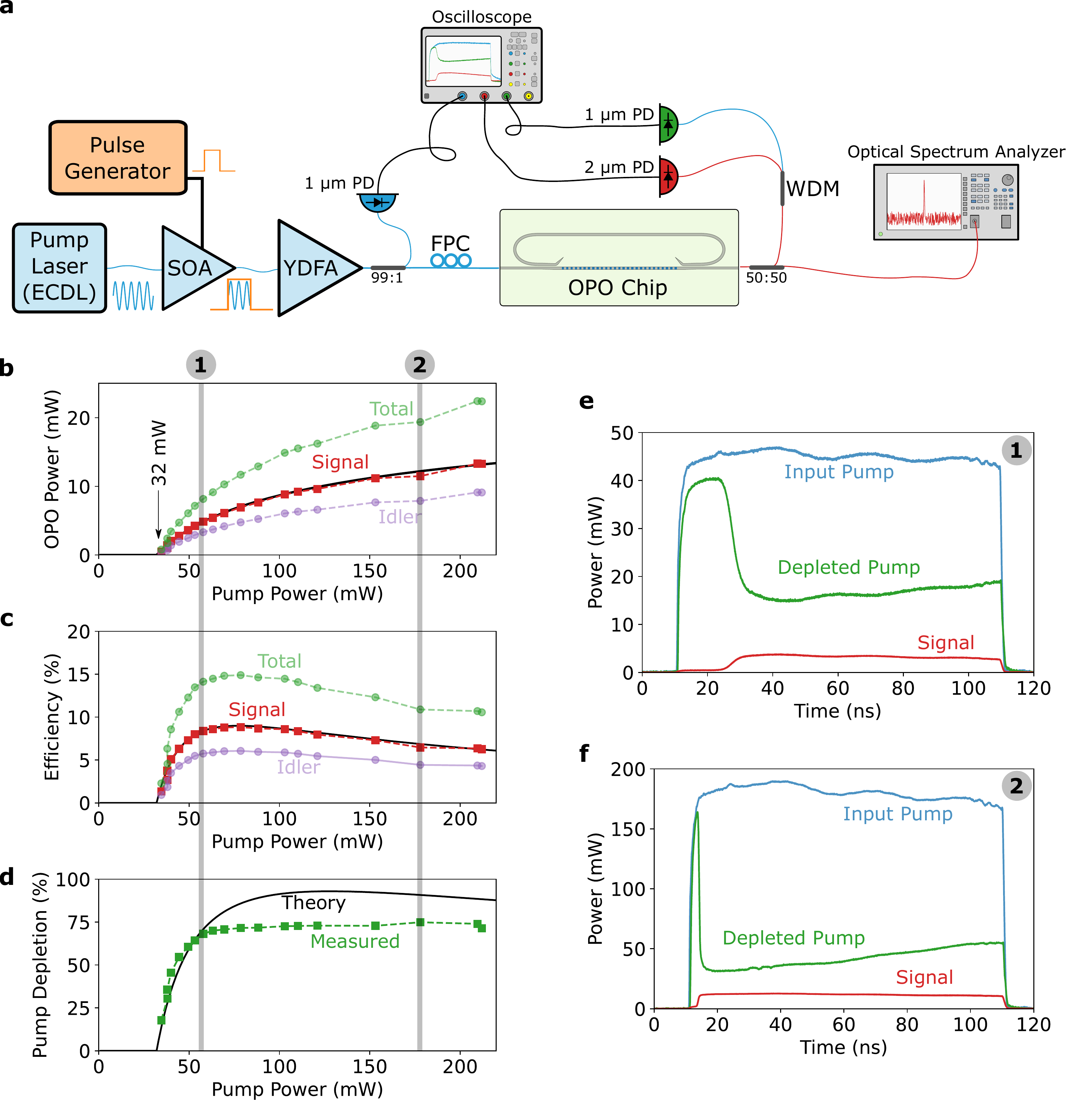}
\caption{
\textbf{Transient and steady-state measurements of on-chip doubly-resonant OPOs.}
\textbf{a}, Measurement setup. We use 100-ns pulses with a 10-kHz repetition rate to decrease the average power while keeping the peak power above the OPO threshold. ECDL, external cavity diode laser; SOA, semiconductor optical amplifier; YDFA, ytterbium doped fiber amplifier; FPC, fiber polarization controller; OPO, optical parametric oscillator; PD, photodetector; WDM, wavelength division multiplexer.
\textbf{b}, On-chip output power versus pump power for a signal wavelength of 1950 nm and a pump wavelength of 1050 nm, the idler and total power are estimated from the signal (see Methods).
\textbf{c}, Different measured on-chip efficiencies.
\textbf{d}, Measured and expected pump depletion levels representing the conversion efficiency within the OPO.
\textbf{e, f}, Measured pump and signal traces at two different power levels as indicated by the shaded gray regions in \textbf{b,c,d}.
}
\end{figure*}

To study the transient and steady-state behaviors of the nanophotonic OPOs we use pulses that are much longer than the cavity lifetime of the OPOs. This arrangement also allows us to use low average powers incident on the chip while maintaining high peak powers. The experimental setup is shown in Fig. 2a, which is described in detail in Methods.

\begin{figure*}
\centering
\includegraphics[width=0.95\linewidth]{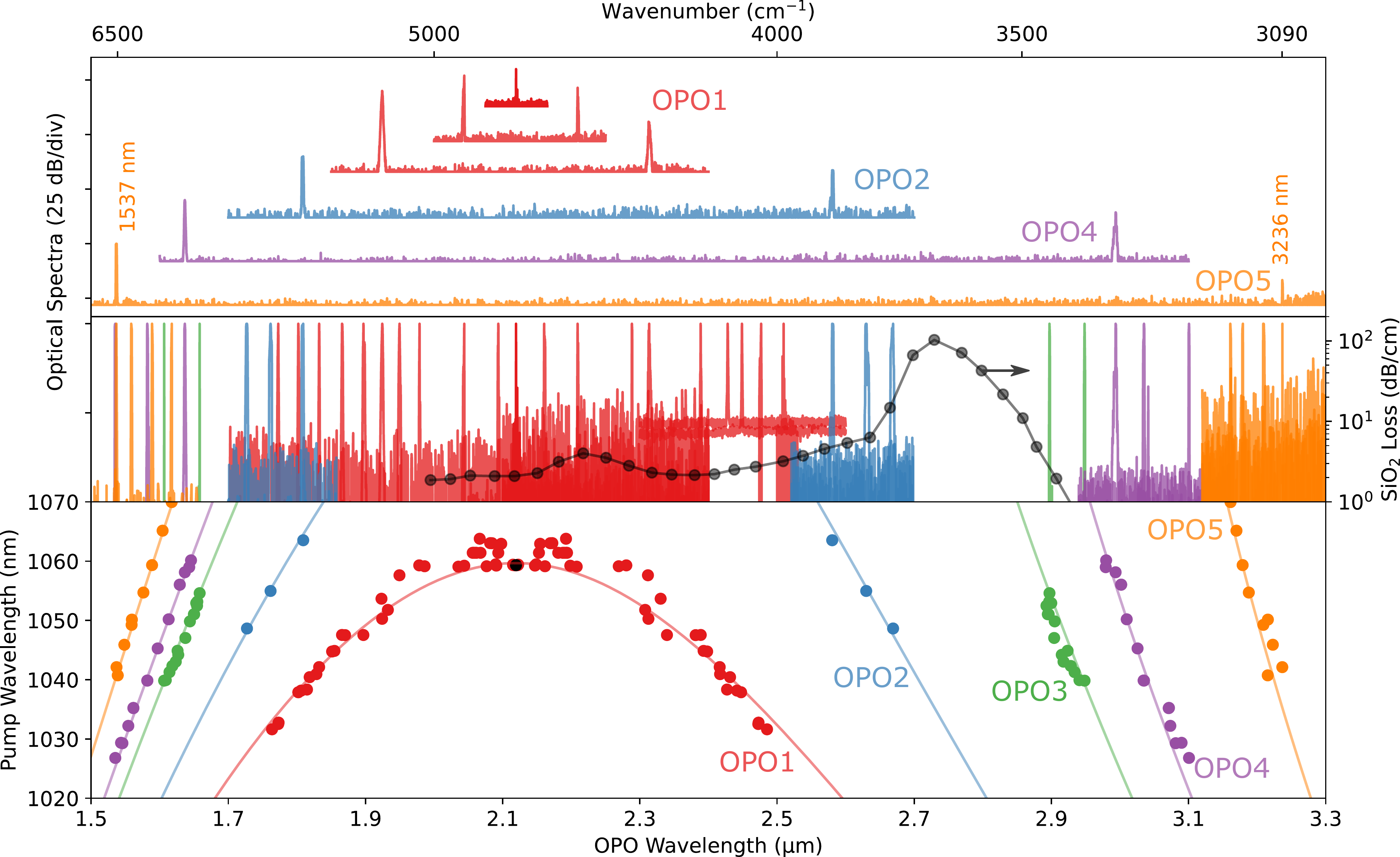}
\caption{
\textbf{Wavelength tuning of nanophotonic OPOs.}
Top panel shows examples of output spectra for a few OPOs on the same chip exhibiting an octave-wide tuning range. Each color represents a different OPO. 
Middle panel shows many more output spectra from the same OPOs.
Bottom panel includes all measured data (colored dots) with the corresponding pump wavelength on the vertical axes along with the theoretical tuning curves (solid lines).
}
\end{figure*}

The measured OPO on-chip signal power at \textasciitilde 1950 nm is shown in Fig. 2b as a function of on-chip pump power (at 1050 nm).  Only the signal (red squares) is measured, as the photodetector is not sensitive to the idler wave near 2275 nm. The idler power (purple circles) is estimated from the output coupler response (see Methods). The solid black line is a fit based on a theoretical expression with an oscillation threshold of \textasciitilde 32 mW. Figure 2c shows the on-chip conversion efficiency, which has a maximum value of \textasciitilde 9 \% for the signal, and up to \textasciitilde 15\% when including the idler. This efficiency is limited by the escape efficiency of the OPO (see Methods) which is currently low for the idler, and can be enhanced significantly with different coupler designs. Pump depletion characterizes the efficiency with which pump photons are converted into signal and idler photons inside the OPO (see Methods). As shown in Fig. 2d, \textasciitilde 75\% is observed, highlighting the potential of nanophotonic OPOs as extremely efficient wavelength conversion devices. These large pump depletion levels are also readily apparent from the oscilloscope traces shown in Fig. 2e,f.


Figure 3 shows the spectral tuning range of five OPOs fabricated on the same chip. The top panel of Fig. 3 shows few spectra of the signal and idler emission of the OPOs. This includes an OPO (OPO1 - red traces) that can operate at degeneracy (top trace), and an OPO (OPO5 - orange traces) that can achieve signal and idler wavelengths separated by more than an octave, and with an idler wavelength well into the mid-infrared.

More spectra from the same OPOs are shown in the middle panel of Fig. 3, demonstrating dense coverage over the entire spectral range, except for a band around \textasciitilde 2.8 $\mu$m where the SiO$_2$ buffer layer exhibits an absorption peak \cite{soref2006}. The tuning parameter in all these cases was the pump wavelength as illustrated in the vertical axis of the bottom panel. Note that OPO1 can be tuned between 1.76 $\mu$m and 2.51 $\mu$m (over 750 nm) by varying the pump wavelength by only 30 nm, corresponding to a tuning magnification factor of \textasciitilde 12 in frequency units. OPO1 can also operate at degeneracy by using a 1060-nm pump as shown in the topmost trace of the top panel, corresponding to the black dot in the bottom panel.

\begin{figure*}[t!]
\centering
\includegraphics[width=0.8\linewidth]{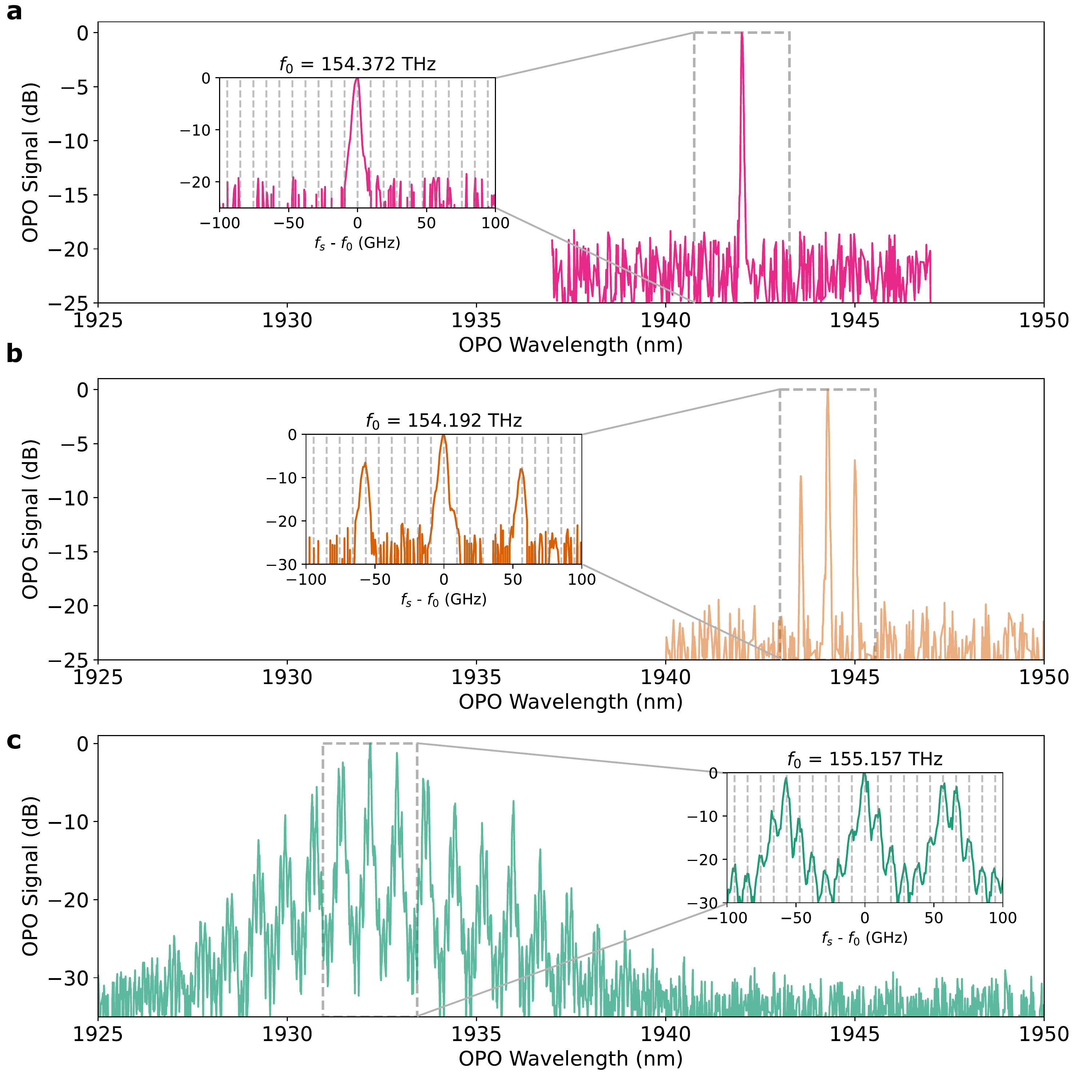}
\caption{
\textbf{Spectral structures of free-running OPOs.}
The generated spectrum of our OPOs can vary from (\textbf{a}) single mode emission, to (\textbf{b}) emission in a few modes separated by several FSRs, to (\textbf{c}) emission in several mode clusters.
Insets show close up of spectra with dashed vertical lines separated by the resonator's FSR which is approximately equal to the 10-GHz OSA resolution bandwidth.
}
\end{figure*}

By tuning the pump power level, the OPOs can operate with a single mode, few modes, or multiple mode clusters, with examples shown in Fig. 4. Closer to threshold the OPOs can oscillate in a single spectral mode as shown in Fig. 4a. As the pump power is increased, oscillation in a few modes can occur as shown in Fig. 4b. Multiple mode clusters appear several times above threshold as shown in Fig. 4c. The multimode behavior is due to the parametric gain-bandwidth being larger than 1 THz, so a large number of modes (\textasciitilde 10 GHz FSR) experience gain. At the same time, waveguide dispersion causes a difference in FSR between signal and idler wavelengths which produces cluster effects in doubly-resonant OPOs well above the threshold \cite{eckardt1991}. Further dispersion and cavity engineering can be employed for either suppressing the multimode effects or tailoring it towards generation of frequency combs \cite{PRL_MOSCA}.


\begin{figure}[t!]
\centering
\includegraphics[width=0.99\linewidth]{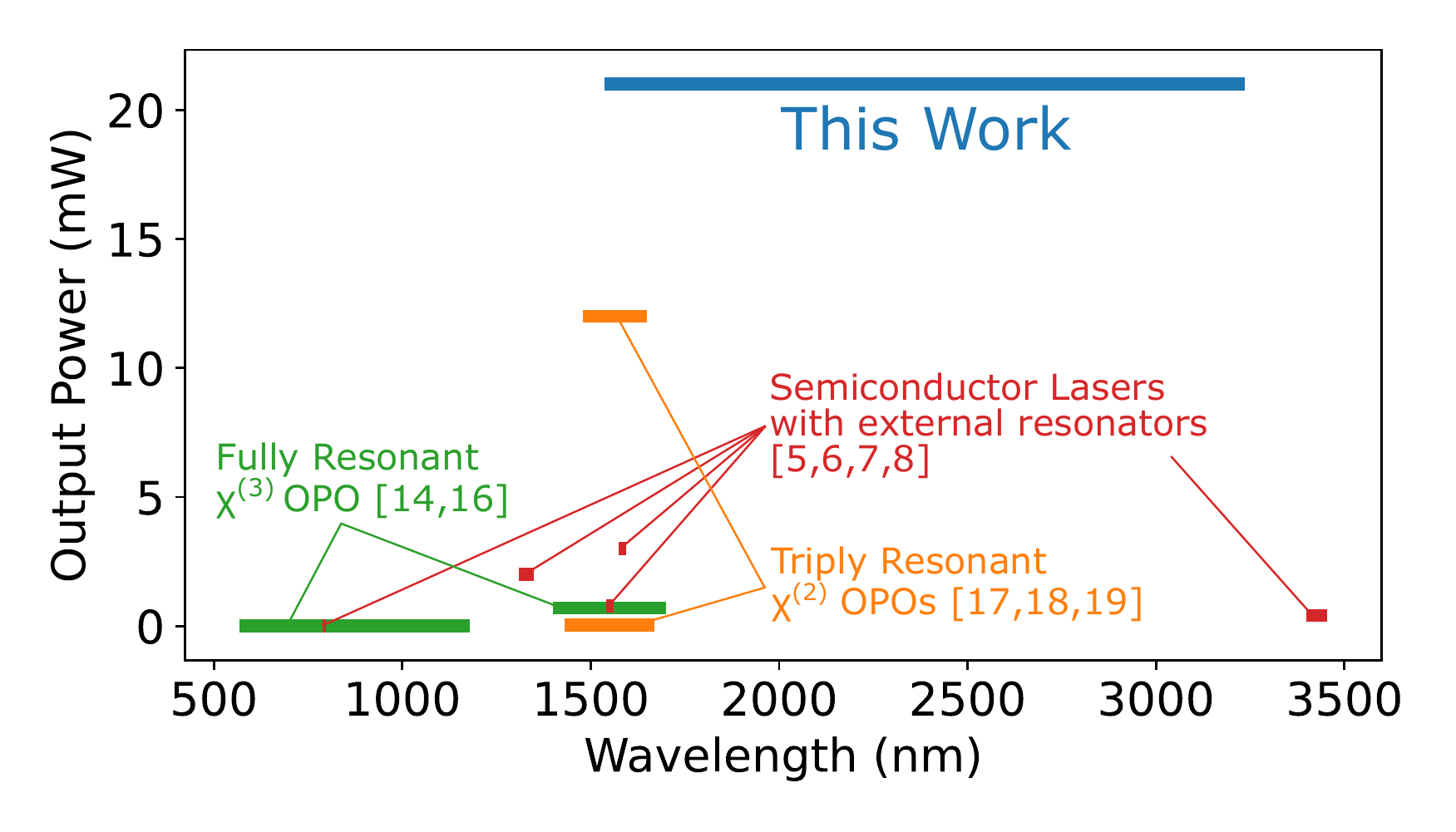}
\caption{
Comparison of the on-chip output power and wavelength coverage of our on-chip OPOs with other integrated tunable sources. The power level is the sum of signal and idler for all OPOs.
}
\end{figure}

Figure 5 shows the tuning range and power level of our OPO chip alongside previously reported tunable sources in nanophotonics. Such significant performance is enabled by the novel non-resonant-pump OPO design combined with dispersion-engineered, quasi-phase matched, directly-etched waveguides (see Methods).


\section*{Discussion}

Our results show that ultra-widely tunable infrared sources can be implemented on the thin-film lithium niobate platform, adding to the increasingly large set of functionalities available in this platform \cite{zhu2021a}, and complementing the recent demonstration of tunable near-infrared DBR lasers \cite{verrinder2022}.

The threshold and required pump tuning range of our OPOs are within the reach of low-cost near-infrared laser diodes. Additional engineering of the cavity design, waveguide dispersion, and quasi-phase matching can be utilized for tailoring the operation towards a multitude of applications. For instance, the threshold of the OPOs can be substantially reduced by utilizing a separate resonator for the pump without sacrificing the conversion efficiency and tunability (unlike a triply-resonant design). 

The maximum conversion efficiency of a doubly-resonant OPO is dominated by its escape efficiency, which is related to the ratio of output coupler transmittance to total resonator losses (see Methods). For our device, this ratio is \textasciitilde 9\% at 1950 nm, indicating that the output coupling is small compared to the total losses in the resonator. This could be caused in part by the little transmission of the output coupler, particularly at mid-infrared wavelengths, and in part by intrinsic resonator losses and losses at the input coupler. Fine tuning of the coupler designs and reducing the cavity loss can lead to substantial improvement of efficiency. We used adiabatic couplers in this work since they provide a simple means to approximately achieve our requirements of high signal and idler coupling together with low pump coupling. However, the input coupler should, ideally, have 100\% coupling at signal and idler frequencies since any transmission in this coupler behaves as additional resonator loss, leading to higher thresholds and lower efficiencies. Simultaneously, the input coupler should provide very low coupling at the pump wavelength, since any coupling just leaks pump power into the unused port, and also provides an undesired feedback path for the pump. These characteristics may be achievable through more advanced coupler designs, for instance, those obtained by inverse design methods \cite{molesky2018}. 

The tuning range of a single OPO can be further enhanced by implementing multiple poling periods on the same OPO. Moreover, since the wavelength coverage of the OPO appears to be limited by the loss of the SiO$_2$ buffer layer, a similar design with a different buffer layer material can allow operation towards the entire lithium niobate transparency window \cite{mishra2021}. The OPO design we demonstrate here can also be readily applied to other emerging nonlinear photonic platforms with transparency windows deeper into the mid-infrared \cite{becheker2022}.

The measurements presented in Fig. 3 only exploit the dependence of the output wavelength on pump wavelength. Two additional degrees of freedom are the temperature and the resonator's free spectral range (which could be varied, for instance, by electro-optic modulation of the resonator's feedback arm). These three variables combined can facilitate precise and fast tuning of the output wavelengths over a much broader spectral range \cite{eckardt1991}, especially when an integrated pump laser is used.

Singly-resonant OPOs offer even smoother tunability and stability characteristics at the expense of higher threshold powers. While pure singly resonant behavior can be obtained by changing the coupler response so that only the signal or idler resonates, we note that the transition between doubly- and singly-resonant designs is smooth \cite{yang1993a} and we have evidence that our OPOs can operate in this regime (see Extended Data Fig. \ref{XtraDROSRO}). This could enable fast and ultrabroad wavelength synthesis on-chip with potential mode-hop free operation.

In summary, we have demonstrated on-chip doubly-resonant OPOs that can be tuned over an octave up to 3.25 $\mu$m. Our OPOs are based on a novel on-chip doubly-resonant design that avoids many of the challenges present in triply-resonant configurations and linear cavity oscillators, and can be easily extended to singly-resonant configurations. Further dispersion engineering may lead to femtosecond synchronously-pumped OPOs in nanophotonics and the numerous applications they unlock \cite{kobayashi2015}.


\section*{Materials and Methods}

\subsection*{Device Design}
We use adiabatic couplers to create the wavelength selective cavity (see Extended Data Fig. \ref{XtraDesign}a). The input and output couplers are identical and are designed so that signal and idler wavelengths ($\lambda > 1.8 \; \mu$m) have large coupling factors ($> 80 \%$) while pump wavelengths near 1 $\mu$m are only slightly coupled ($< 10 \% $). The residual coupling of the pump leads to round-trip feedback factors of less than $1 \%$ that produce negligible modulations of the pump intensity as a function of frequency, allowing continuous tuning of the pump wavelength.

When designing a tunable OPO, it is desirable to have a large tuning slope so a small change in pump wavelength produces large changes in the output wavelengths. At the same time, a small gain-bandwidth is preferable to limit the number of resonator modes experiencing gain. To achieve a balance between these two behaviors, we engineer the dispersion of the waveguide using its geometry, resulting in 2.5-$\mu$m-wide waveguides on a 700-nm-thick lithium niobate layer and 250 nm of etching depth. The mode profile for a set of representative wavelengths are shown in Extended Data Fig. \ref{XtraDesign}b, illustrating that the modal overlap remains substantial despite the large frequency difference.

\subsection*{Device Fabrication}
We fabricate our devices using a commercial wafer (NANOLN) with an x-cut, 700-nm-thick MgO-doped lithium niobate layer and a SiO$_2$ buffer layer. We provide quasi-phase matching in a 5-mm-long region through periodic poling (inset of Fig. 1b shows a second-harmonic microscope image of a typical poled section). The waveguides are patterned by e-beam lithography and dry etched with Ar$^+$ plasma to a depth of 250 nm. All the OPOs have the same waveguide geometry obtained from dispersion engineering, with 2.3-$\mu$m-wide input and output waveguides that taper (through the adiabatic couplers) to 2.5-$\mu$m-wide waveguides inside the resonator. To maximize the spectral range covered on a single chip, we fabricated OPOs with poling periods ranging from 5.55 $\mu$m to 5.7 $\mu$m in 10-nm steps.  We include a straight waveguide next to each OPO for calibration and quasi-phase matching verification (colored purple in Fig. 1b).

\subsection*{Device Characterization}
We characterize our OPOs using the experimental setup shown in Fig. 2a, that consists of a tunable CW 1-$\mu$m laser amplified by a semiconductor optical amplifier (SOA) which is modulated to generate 100-ns-long (full-width-half-maximum) pulses with 10-kHz repetition rate. These pulses are further amplified by an ytterbium doped fiber amplifier (YDFA) and coupled into the chip using a single-mode 1-$\mu$m lensed fiber (\textasciitilde 10 dB coupling loss). The OPO output is collected either by a 2-$\mu$m lensed fiber, or a cleaved InF$_3$ fiber, and sent to an optical spectrum analyzer (OSA) or to an InAsSb detector connected to an oscilloscope. A wavelength division multiplexer (WDM) allows us to monitor the depleted pump and signal output simultaneously.

To estimate propagation losses in our waveguides, we have fabricated chips with arrays of critically coupled resonators and extracted quality factors \textasciitilde $6\times 10^5$, which translate to losses below 0.3 dB/cm for waveguides without poling. Detailed inspection of the periodically poled waveguide inside the resonator reveals periodic roughness of the waveguide sidewalls, likely from the polarization-dependent etch rate of lithium niobate. More studies are needed to improve the resonator quality factor. 

Input and output coupling losses are estimated from several measurements on straight waveguides. Comparing transmission of straight waveguides to that of OPOs allows to estimate a total loss factor of 0.929 per coupler at the pump wavelength, reasonably close to the simulated value of 0.95 (Extended Data Fig. \ref{XtraDesign}a). The plots of Fig.2b-d are obtained from oscilloscope traces like those in Fig.2e,f by first converting voltage to power, integrating them to find the energy, and then dividing by the 100-ns pulse width to obtain the average peak power.

\subsection*{Efficiency and Idler Power Estimation}

The efficiency of an OPO ($\eta$) can be written as the product of two efficiencies, $\eta = \eta_0 \eta_\mathrm{escape}$. The internal efficiency ($\eta_0$) measures how efficiently pump photons are converted into signal and idler photons, while the escape efficiency ($\eta_\mathrm{escape}$) measures the fraction of the generated signal and idler photons available at the output of the OPO. The difference between the pump power at the beginning and end of the gain section is $\Delta P_p = P_p(0) - P_p(L_g)$. The internal efficiency is just the pump depletion $\eta_0 = \Delta P_p / P_p(0)$ shown in Fig 2d.

The escape efficiency is given by 
\begin{align*}
    \eta_e(\omega) = \frac{\omega}{\omega_p} \frac{T(\omega)}{1-L(\omega)},
\end{align*}
where $T(\omega)$ is the power transmission coefficient of the output coupler, while $L(\omega)$ is the roundtrip power loss factor of the resonator. Since the output power can be calculated from the efficiency as $P_\mathrm{out}(\omega) = \eta_0 \eta_e(\omega) P_p(0)$, the idler power can be calculated from the signal power as:
\begin{align*}
    P_\mathrm{out}(\omega_i) \approx \frac{\eta_e(\omega_i)}{\eta_e(\omega_s)} P_\mathrm{out}(\omega_s).
\end{align*}

\textbf{Funding.}  The authors gratefully acknowledge support from ARO grant no. W911NF-18-1-0285, NSF grant no. 1846273 and 1918549, AFOSR award FA9550-20-1-0040. 

\textbf{Acknowledgment.}
The device nanofabrication was performed at the Kavli Nanoscience Institute (KNI) at Caltech. This work was supported by a NASA Space Technology Graduate Research Opportunities Award. Part of this research was carried out at the Jet Propulsion Laboratory, California Institute of Technology, under a contract with the National Aeronautics and Space Administration. The authors thank NTT Research for their financial and technical support.

\textbf{Disclosures.} L.L, R.M.B. and A.M: US patent 11,226,538 (P). The remaining authors declare no conflicts of interest.

\textbf{Data availability.} Data underlying the results presented in this papers may be obtained from the authors upon reasonable request.

\newpage


\bibliography{references.bib}

\renewcommand{\figurename}{Extended Data Fig.}
\renewcommand{\tablename}{Extended Data Table}
\setcounter{figure}{0} 
\setcounter{table}{0}

\begin{figure*}[t!]
\centering
\includegraphics[width=0.99\linewidth]{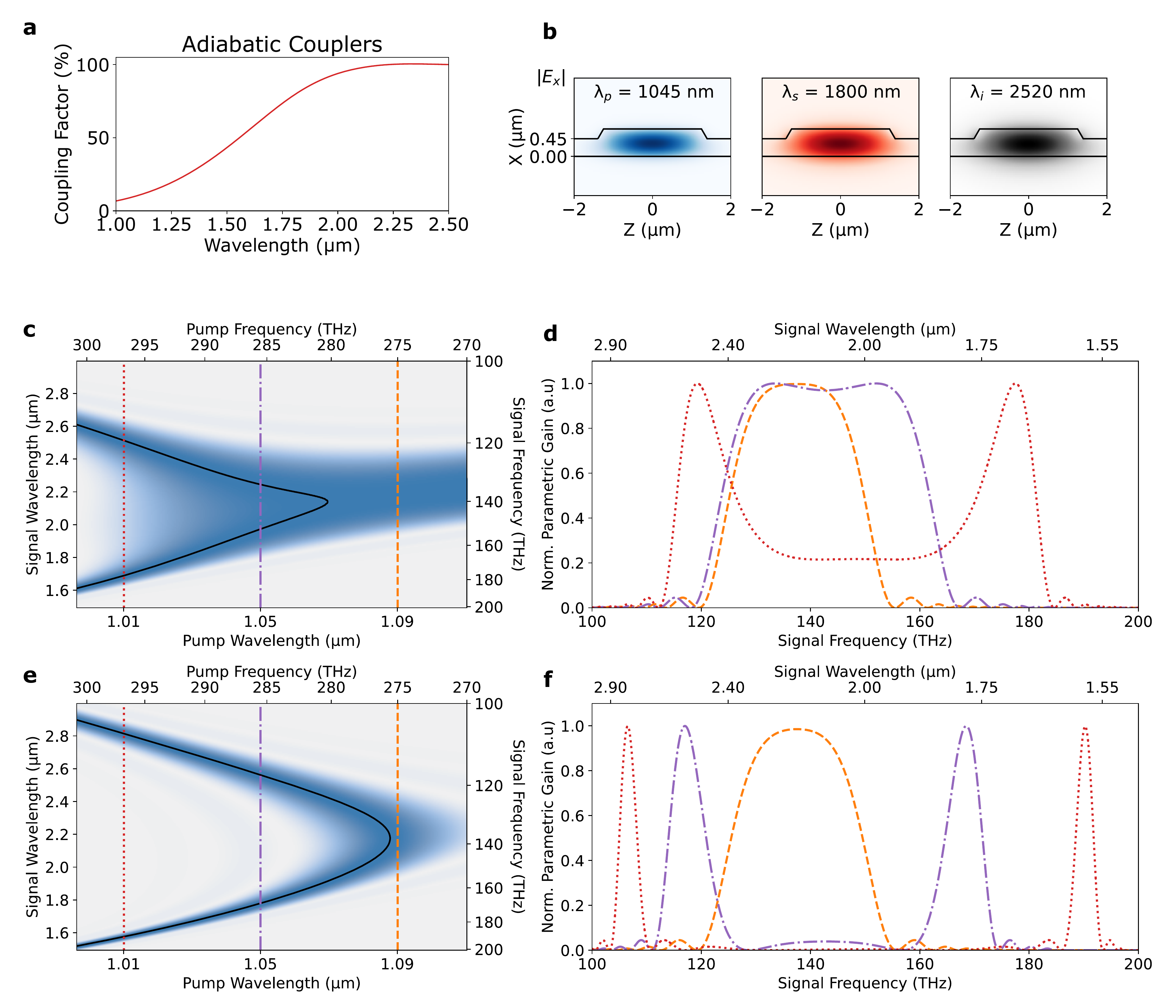}
\caption{
\textbf{Doubly-resonant OPO Design.}
\textbf{a}, Simulated coupling factor of adiabatic couplers used to create the OPO resonator.
\textbf{b}, Simulated mode profiles for a representative set of pump, signal, and idler wavelengths, illustrating their similarities despite the large frequency
difference leading to a substantial mode overlap over wide bandwidth.
\textbf{c-f}, Two examples of OPOs with different geometries, showing the
effect of dispersion engineering on the tuning curves. 
\textbf{c,d} Correspond to an OPO that could
be pumped with femtosecond pulses for frequency comb generation in the mid-infrared. 
\textbf{e,f} Correspond to the OPOs described in the main text, exhibiting a smooth tuning characteristic.
\textbf{d,f} Parametric gain as a function of signal (and idler) frequency for fixed pump
wavelengths indicated by the vertical lines in \textbf{c,e}.
}
\label{XtraDesign}
\end{figure*}

\begin{figure*}[t!]
\centering
\includegraphics[width=0.99\linewidth]{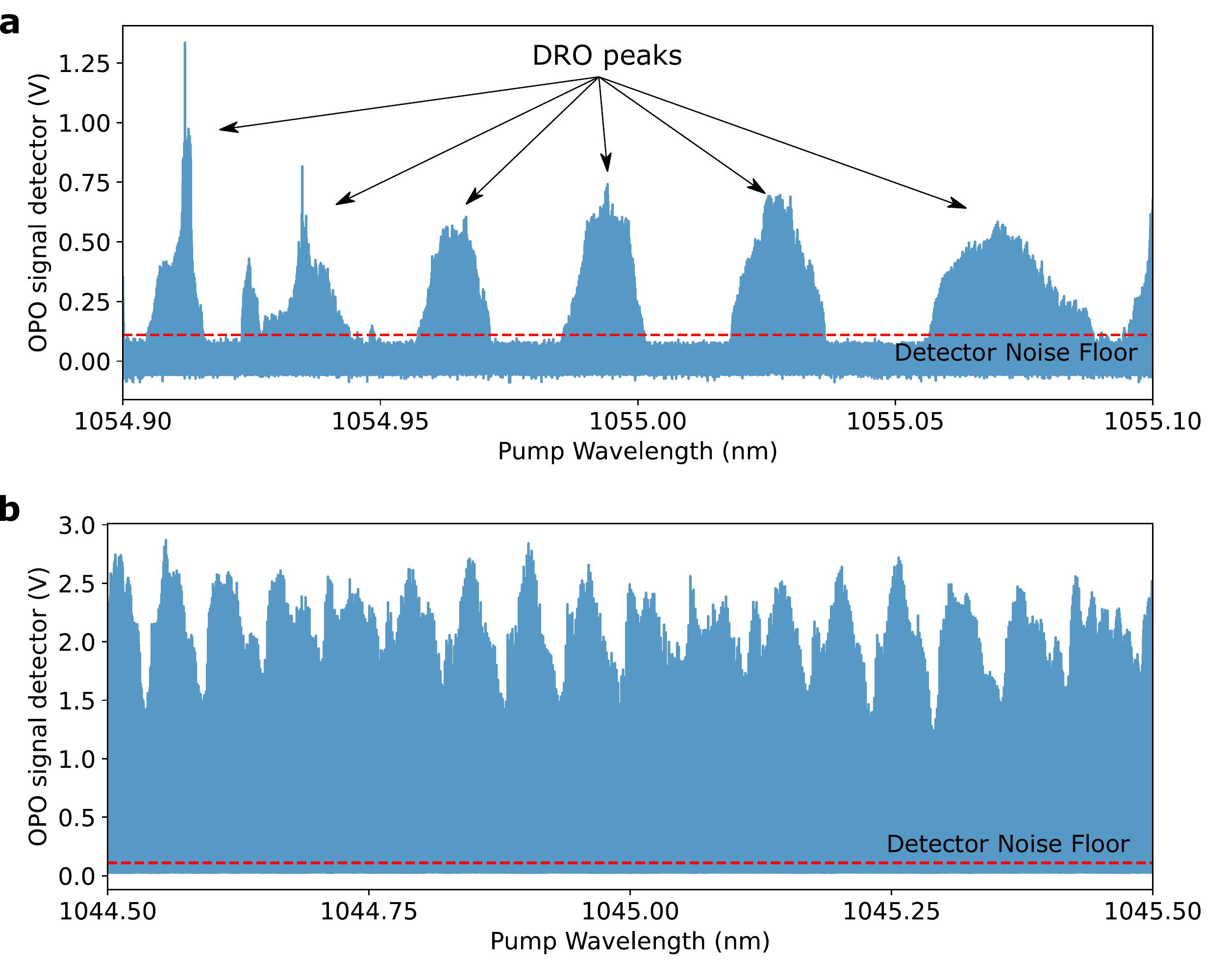}
\caption{
\textbf{Doubly-resonant and singly resonant regimes.}
\textbf{a}, Detected OPO signal for low pump power showing isolated oscillation peaks. This phenomenon, known as cluster effects in doubly-resonant OPOs (DROs), is due to the difference between the free spectral range at signal and idler wavelengths produced by waveguide dispersion.
\textbf{b}, At larger pump powers the OPO oscillates for any pump wavelength. This is because away from a doubly-resonant cluster, the OPO operates closer to the singly resonant regime, with a strongly resonant idler and a weakly resonant signal that is free to adjust itself to a frequency $\omega_s = \omega_p - \omega_i$. In both, \textbf{a} and \textbf{b}, signal power variations are due to a combination of threshold variations and wavelength dependent pump laser power. 
}
\label{XtraDROSRO}
\end{figure*}

\end{document}